\documentclass[preprint2]{aastex}

\shorttitle{Initial Energy of SNRs}
\shortauthors{J. W. Xu et al.}

\begin{document}

\title{A Unique Method to Determine SNe Initial Explosion Energy}

\author{Jian-Wen Xu\altaffilmark{1} and Hui-Rong Zhang\altaffilmark{2}}
\affil{Key Laboratory of Frontiers in Theoretical Physics, Institute
of Theoretical Physics, Chinese Academy of Sciences, Beijing 100080,
China}

\altaffiltext{1}{Postdoctor, Institute of Theoretical Physics,
Chinese Academy of Sciences, Beijing 100080, China.}
\altaffiltext{2}{Postdoctor, Institute of Theoretical Physics,
Chinese Academy of Sciences, Beijing 100080, China.}
\email{xjw@itp.ac.cn}

\begin{abstract}
There are several different methods to determine the individual
supernovae (SNe) initial explosion energy, here we derive the
average or typical explosion energy of shell-type supernova remnants
(SNRs) in a particular way. By solving a group of equations
pertaining to shell-type SNRs at the same stage we obtained some
physical parameters, e.g. the distance ($d$), evolved age ($t$),
etc.. Assuming series of different SN initial explosion energies
ranging from $10^{48}$~ergs to $10^{53}$~ergs, we derived series of
distance and age parameters with which compared already known ones.
Thus the most likely value of the SNe initial explosion energy is
obtained when the deviation is least, which equals to about
$10^{51}$~ergs, in good agreement with the undertook value.
\end{abstract}


\section{Introduction}

We have already known the initial kinetic energy ($E_0$) of lots of
Galactic SNRs through various estimate methods
(table~\ref{tabdistr}). These SNRs embody shell-type and
composite-type ones. Fig.~\ref{distrib} denotes the number
distribution of 44 such remnants. It seems that they concentrate to
about $10^{50} \sim 10^{51}$~ergs. Therefore we can significantly
take the average initial energy ($E_0$) as a typical SNR physical
parameter which is also able to be determined by our unique means.

Our mathematical method to derive SNe initial energy is rather
different from others. One first method of others, for example, by
$E = \frac{1}{2} M_{su}\upsilon^2$, $M_{su}$ is the swept-up mass of
remnant shell expanding into interstellar media (ISM), $\upsilon$ is
the velocity of shock wave of remnant, the initial explosion energy
of SNR G180.0$-$1.7 is thus obtained (Braun et al. 1989). Sun et al.
(1999) calculating their detected ASCA data plus ROSAT data has
derived the initial energy of G327.1$-$1.1 as the fitting result.
After knowing the SNR G299.2$-$2.9 radius ($R$) value, the particle
density ($n_0$) and the age ($t$), Slane et al. (1996) get $E_0$
value by $E_0 \approx 340 R^5 n_0 t^{-2} \times 10^{51}$~ergs. Bamba
et al. (2001) have got the $E_0$ value by assuming a thin thermal
NEI plasma model plus standard Sedov model. And so on. All of them
obtain the initial energy of a individual SNR through physical
means. Here in the paper we get SNe average initial explosion energy
by statistical method.

Galactic supernova remnants are classified into three types:
shell-type, Plerion-type and composite-type. Our work here merely
include the shell-type remnants. Moreover, shell type SNRs usually
have four evolution stages: the free expansion phase, the Sedov or
adiabatic phase, the radiative or snowplough phase and the
dissipation phase. Nearly all the observed SNRs are in the adiabatic
phase, or in the 3rd. And almost none is detected in the 1st and 4th
phases.

In the paper, numerical analysis for the most likely value of the SN
initial energy is described in section~\ref{theory}, of which they
are made separately at adiabatic-phase and radiative-phase for
shell-type SNRs, and both made comparison by distances and ages. In
the last section we discuss and

\clearpage

\begin{deluxetable}{llllll}
\tablewidth{0pt} \tablecaption{List of the initial explosion energy
$E_0$ of 44 Galactic shell-type or composite-type SNRs of which
their value was somewhat well determined by various methods.
\label{tabdistr}}
\startdata \hline\noalign{\smallskip} \hline\noalign{\smallskip}
 Source & $E_0\times10^{51}$ & Ref.   & Source   & $E_0\times10^{51}$ & Ref.\\
    $-$        & ergs        &  $-$   &  $-$     &  ergs              & $-$\\
\hline\noalign{\smallskip}
   G0.0$+$0.0  & $\ge$40\tablenotemark{a}& KF96  & G180.0$-$1.7 & 0.24  & BGL89\\
   G4.5$+$6.8  & 0.4/0.5\tablenotemark{b}& BKV05 & G184.6$-$5.8 & 0.015 & GAT04a\\
  G18.8$+$0.3  & 0.01        & D99    & G261.9$+$5.5  & 0.29          & CD80  \\
  G18.9$-$1.1  & 0.08$-$0.18 & H04    & G263.9$-$3.3  & 1$-$2         & GAT04a\\
  G24.7$+$0.6  & 0.066       & L89    & G290.1$-$0.8  & 0.8           & S02   \\
  G28.6$-$0.1  & 0.9         & BUK01  & G291.0$-$0.1  & 0.25?         & HHS98 \\
  G29.7$-$0.3  & 2           & HCG03  & G292.0$+$1.8  & 0.18          & GAT04b\\
  G31.9$+$0.0  & 0.3$-$1.4\tablenotemark{c}& C05 & G292.2$-$0.5 & 1.0 & GAT04b\\
  G39.7$-$2.0  & 1           & MS96   & G296.1$-$0.5  & 0.23          & HM94  \\
  G41.1$-$0.3  & 1.2         & SDP05  & G296.5$+$10.0 & 0.2$-$0.6     & GAT04b\\
  G74.0$-$8.5  & 0.24        & BGL89  & G299.2$-$2.9  & 0.12          & SVH96 \\
  G82.2$+$5.3  & 0.17        & M04    & G312.4$-$0.4  & 0.6           & CB99  \\
  G93.3$+$6.9  & 0.39        & LRR99  & G315.4$-$2.3  & 0.66          & GAT04b\\
  G94.0$+$1.0  & $\ge$0.27   & F05    & G320.4$-$1.2  & 1$-$2         & GAT04b\\
 G106.3$+$2.7  & $>$0.07     & KUP01  & G326.3$-$1.8  & 1.0           & GAT04b\\
 G109.1$-$1.0  & 1$-$10      & GAT04a & G327.1$-$1.1  & 0.23          & SWC99 \\
 G111.7$-$2.1  & 2$-$3       & V06    & G327.6$+$14.6 & 1.0           & GAT04b\\
 G116.9$+$0.2  & 0.1         & CHP97  & G347.3$-$0.5  & 1.0           & MTT05 \\
 G119.5$+$10.2 & 0.03        & GAT04a & G349.7$+$0.2  & 0.5           & L05   \\
 G120.1$+$1.4  & 1.16        & WHB05  & G352.7$-$0.1  & 0.2           & K98a  \\
 G126.2$+$1.6  & $>$7        & B05    & G357.7$+$0.3  & 0.11          & L89   \\
 G132.7$+$1.3  & 0.31        & GAT04a & G359.0$-$0.9  & 0.21          & L89   \\
\enddata
\tablenotetext{a}{Notes: In this case we just make use of the
minimum value for our purpose.} \tablenotetext{b}{Notes: In the case
we take the average value.} \tablenotetext{c}{Notes: The same as
above.}
\end{deluxetable}

\clearpage

summarize our results.

\section{Numerical Analysis}\label{theory}

\subsection{At adiabatic phase}\label{sedov}

Let us list the group of equations for shell-type supernova remnants
at the second stage as follow (Wang \& Seward 1984, Koyama \&
Meguro-Ku 1987, Bignami \& Caraveo 1988, Xu et al. 2005),

\begin{equation}
D_{pc} = 4.3\times 10^{-11} \left( \begin{array}{c}
\frac{E_0}{n_{cm^{-3}}} \end{array} \right) ^{1/5} t_{yr} ^{2/5}
\end{equation}

\begin{eqnarray}
\Sigma(D) = 1.505\times 10^{-19} \frac{S_{1GHz}}{\theta_{arcmin}^2}\nonumber\\
= 2.88\times 10^{-14} D_{pc}^{-3.8} n_{cm^{-3}}^2
\end{eqnarray}

\begin{eqnarray}
\left( \begin{array}{c} \frac{E_0}{10^{51}ergs} \end{array} \right)
= 5.3\times10^{-7} n_{cm^{-3}}^{1.12} \upsilon _{Km~s^{-1}}^{1.4}\nonumber\\
\times \left(\begin{array}{c} \frac{D_{pc}}{2}
\end{array} \right)^{3.12}
\end{eqnarray}

Here, $D_{pc}$ is the SNR diameter in unit of pc, $t_{yr}$ is the
remnant age in year, $n$ is the ISM electron density in $cm^{-3}$,
$S_{1GHz}$ is the detected fluxes of an SNR in Jy at 1~GHz,
$\theta_{arcmin}$ is the observational angle in arcmin, $\upsilon =
\frac{dD}{dt}$ is the velocity of shock waves in km~$s^{-1}$. And we
know $tan \left(\begin{array}{c} \frac{\theta_{arcmin}}{2}
\end{array} \right) = \frac{D_{pc}}{2d_{pc}}$, where, $d_{pc}$ is
the distance to a remnant in pc.

From these parameters above, the fluxes $S_{1GHz}$ and observational
angle $\theta_{arcmin}$ is the detected value for each SNR
(table~\ref{tab2nd}) of which we regard these remnants evolving at
the Sedov-phase since their diameter less than 36~pc. And the SNRs
explosion energy $E_0 = \xi \times E_{48}$ $(\xi = 1, 2, 3, ...,
10^5)$ is an assumed value. But the diameter $D_{pc}$ (and distance
$d_{pc}$), age $t_{yr}$, velocity $\upsilon_{km s^{-1}}$ and
electron density $n_{cm^{-3}}$ is unknown and to be derived.
Parameters of the distance $d_{pc} = d_{our}(\xi,i)$, and the age
$t_{yr} = t_{our}(\xi,i)$ ($\xi = 1, 2, 3, ..., 10^5$, i = 1, 2, 3,
..., 37) will be adopted in our paper. But we do nothing with
$\upsilon$ and $n$.

For a certain supernova remnant ($i$) ($i = 1, 2, 3, ..., 37$)
(table~\ref{tab2nd}) and assumed  series of SNe initial explosion
energies $E_0 = \xi \times E_{48} (\xi = 1, 2, 3, ..., 10^5)$ in the
unit of $10^{48}$~ergs, we can obtain the remnant distance
$d_{our}(\xi,i) (= d_{pc})$ and the age $t_{our}(\xi,i) (= t_{yr})$
values by solving the group of equations above. Then we compare them
with the already known parameters $d_{true}(i)$ and $t_{true}(i)$
($i = 1, 2, 3, ..., 37$) listed in table~\ref{tab2nd} in order to
derive the most likely original energy of remnants. Thus the
explosion energy was derived when the deviation in comparison is
least.

The group of equations are not strictly correct as not to be figured
out mathematically, but they are correct enough for us to confirm
the SNe initial energy ($E_0$).

\begin{figure}[!htbp]
\includegraphics[angle=0,scale=.27]{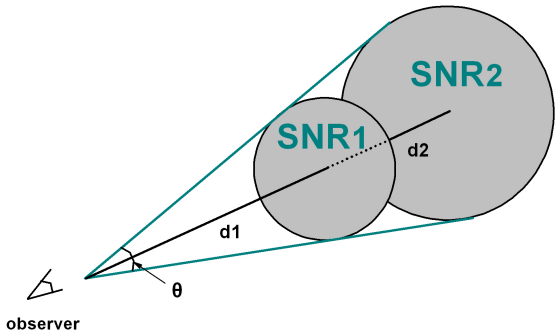}
\caption{The plot shows for both SNRs their radio surface brightness
($\Sigma$) can be equal to each other when $S_{1GHz, 2}/S_{1GHz, 1}
= \theta_2^2/\theta_1^2 = 1$. Here $\theta_i^2 (i = 1, 2)$ is the
visual area of the remnant. Thus we are not able to uniquely derive
the SNR diameter ($D$) and distance ($d$) value. But fortunately
this case will never occur since $\Sigma(D) \sim D_{pc}^{-3.8}$ and
not $\sim D_{pc}^{-2}$. Therefore we can obtain the just remnant
distance ($d$) by solving the group of equations offerred in text.
\label{angl}}
\end{figure}

Figure~\ref{angl} shows that when $S_{1GHz, 2}/S_{1GHz, 1} =
\theta_2^2/\theta_1^2 = 1$ for both SNR1 and SNR2, here $\theta_i^2
(i = 1, 2)$ is the visual area of the remnant, their radio surface
brightness ($\Sigma$) can be the same to each other. Thus the
remnant diameter ($D$) and distance ($d$) value will be uncertain
according to equation (2). But fortunately the true reality will
never take on this case because one can see $\Sigma(D) \sim
D_{pc}^{-3.8}$ from formula (2) and not $\sim D_{pc}^{-2}$.
Therefore we are able to uniquely determine the SNe initial kinetic
energy ($E_0$).

Many of the radio SNRs have more than one published value for
distance and age in table~\ref{tab2nd} and~\ref{tab3rd}. For these,
we either chose the most recent estimates or used an average of the
available estimates, or the most commonly adopted value.

\subsubsection{Comparison by distances}

We can compare these resolved distances ($d_{our}(\xi,i)$) above with
the already known ones

\begin{figure}[!htbp]
\includegraphics[angle=0,scale=.30]{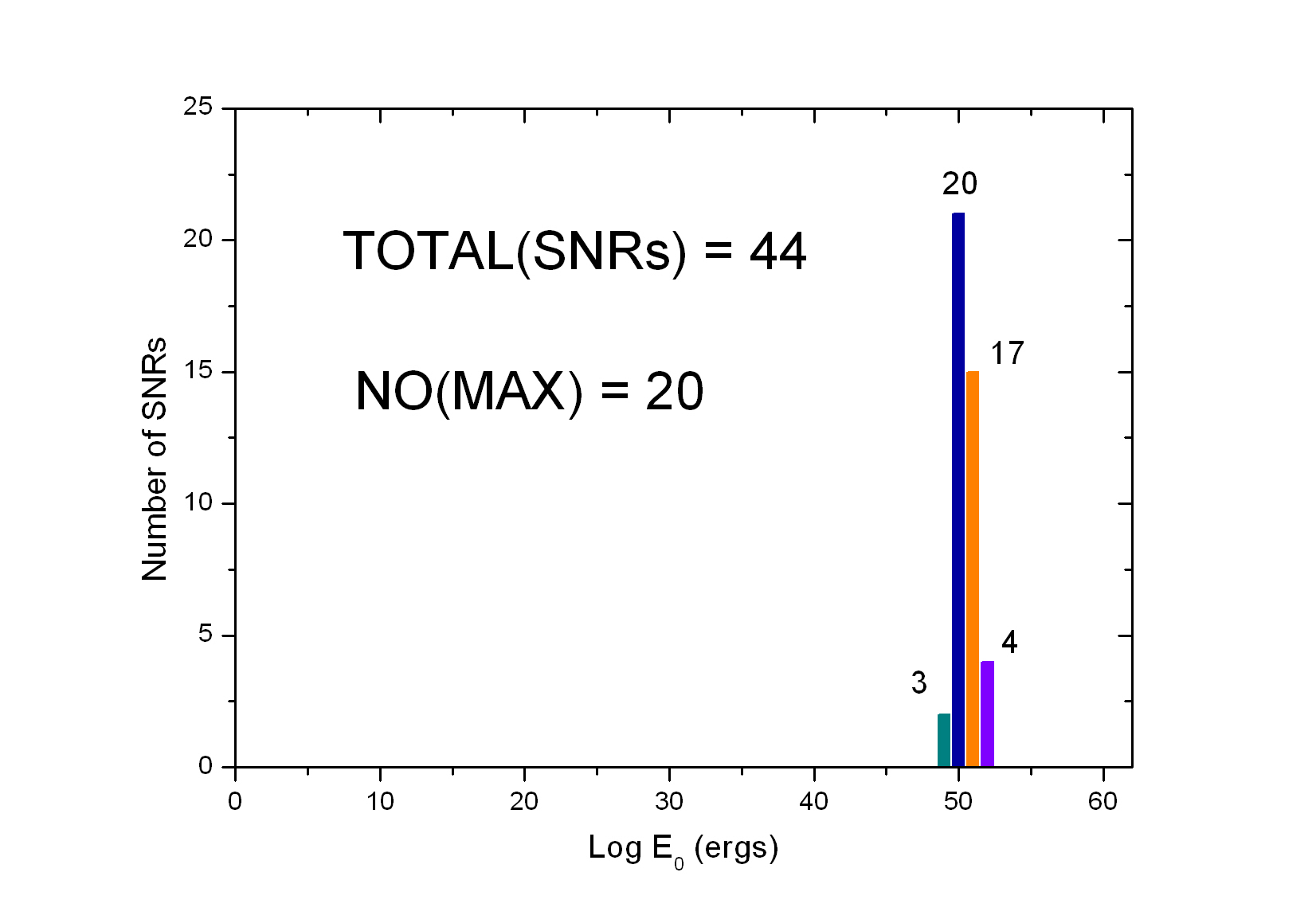}
\caption{the number distribution of Galactic supernova remnants
corresponding to their progenitor initial kinetic energy ($E_0$).
\label{distrib}}
\end{figure}

\begin{deluxetable}{lllllll}
\tablewidth{0pt} \tablecaption{List of the distance ($d$), Age ($t$)
and some other physical parameters of 37 shell-type Galactic SNRs of
which their diameter is less than 36~pc. \label{tab2nd}} \startdata
\hline\noalign{} \hline\noalign{}
 Source & $t_{true}(i)$ & $d_{true}(i)$ & Dia.\tablenotemark{a} & size($\theta$) & $S_{1GHz}$ & Ref.\\
    $-$       &  yr      &   pc        &  pc    &  arcmin   &  Jy     &  $-$        \\
\hline\noalign{}
   G4.5$+$6.8 &  380     &   2900      &  3     &  3        &  19     &  H90, G04a  \\
   G7.7$-$3.7 &  $-$     &   4500      &  29    &  22       &  11     &  M86        \\
  G27.4$+$0.0 &  2700    &   6800      &  8     &  4        &  6      &  C82, G04a  \\
  G31.9$+$0.0 &  4500    &   7200      &  13    &  7x5      &  24     &  CS01       \\
  G32.8$-$0.1 &  $-$     &   7100      &  35    &  17       &  11     &  K98b       \\
  G33.6$+$0.1 &  9000    &   7800      &  23    &  10       &  22     &  S03, SV95, G04a\\
  G39.2$-$0.3 &  1000    &   11000     &  22    &  8x6      &  18     &  C82        \\
  G41.1$-$0.3 &  1400    &   8000      &  8     &  4.5x2.5  &  22     &  C82, B82, C99\\
  G43.3$-$0.2 &  3000    &   10000     &  10    &  4x3      &  38     &  L01        \\
  G53.6$-$2.2 &  15000   &   2800      &  24    &  33x28    &  8      &  S95, G04a  \\
  G73.9$+$0.9 &  10000   &   1300      &  8     &  22?      &  9      &  L89, LLC98 \\
  G74.0$-$8.5 &  14000   &   400       &  23    &  230x160  &  210    &  LGS99, SI01, G04a\\
  G78.2$+$2.1 &  50000   &   1500      &  26    &  60       &  340    &  LLC98, KH91\\
  G84.2$-$0.8 &  11000   &   4500      &  23    &  20x16    &  11     &  MS80, M77, G04a\\
  G89.0$+$4.7 &  19000   &   800       &  24    &  120x90   &  220    &  LA96       \\
  G93.3$+$6.9 &  5000    &   2200      &  15    &  27x20    &  9      &  L99,  G04a \\
  G93.7$-$0.2 &  $-$     &   1500      &  35    &  80       &  65     &  UKB02      \\
 G109.1$-$1.0 &  17000   &   3000      &  24    &  28       &  20     &  FH95, HHv81, G04a\\
 G111.7$-$2.1 &  320     &   3400      &  5     &  5        &  2720   &  TFv01      \\
 G114.3$+$0.3 &  41000   &   700       &  15    &  90x55    &  6      &  MBP02, G04a\\
 G116.5$+$1.1 &  280000  &   1600      &  32    &  80x60    &  11     &  RB81, G04a \\
 G116.9$+$0.2 &  44000   &   1600      &  16    &  34       &  9      &  KH91, G04a \\
 G120.1$+$1.4 &  410     &   2300      &  5     &  8        &  56     &  H90, G04a  \\
 G260.4$-$3.4 &  3400    &   2200      &  35    &  60x50    &  130    &  B94, RG81  \\
 G272.2$-$3.2 &  6000    &   1800      &  8     &  15?      &  0.4    &  D97        \\
 G284.3$-$1.8 &  10000   &   2900      &  20    &  24?      &  11     &  RM86       \\
 G299.2$-$2.9 &  5000    &   500       &  2     &  18x11    &  0.5    &  SVH96      \\
 G309.2$-$0.6 &  2500    &   4000      &  16    &  15x12    &  7      &  RHS01      \\
 G315.4$-$2.3 &  2000    &   2300      &  28    &  42       &  49     &  DSM01, G04a\\
 G327.4$+$0.4 &  $-$     &   4800      &  29    &  21       &  30     &  SKR96, WS88, G04a\\
 G327.6$+$14.6&  980     &   2200      &  19    &  30       &  19     &  SBD84, G04a\\
 G332.4$-$0.4 &  2000    &   3100      &  9     &  10       &  28     &  CDB97, MA86, G04a\\
 G337.2$-$0.7 &  3250    &   15000     &  26    &  6        &  2      &  RHS01      \\
 G337.8$-$0.1 &  $-$     &   12300     &  27    &  9x6      &  18     &  K98b       \\
 G346.6$-$0.2 &  $-$     &   8200      &  19    &  8        &  8      &  K98b, D93  \\
 G349.7$+$0.2 &  14000   &   14800     &  9     &  2.5x2    &  20     &  RM01, G04a \\
 G352.7$-$0.1 &  2200    &   8500      &  17    &  8x6      &  4      &  K98a       \\
\enddata
\tablenotetext{a}{Notes: Diameters were calculated by using the
distances together with the angular sizes in Green (2006)
catalogue.}
\end{deluxetable}

($d_{true}(i)$) listed in table~\ref{tab2nd} by
\begin{eqnarray}
\Phi(\xi, d) = \sum_{i=1}^{37} (d_{our}(\xi,i)-d_{true}(i))^2\nonumber\\
\div \sum_{i=1}^{n} d_{true}^2(i)\nonumber\\
(\xi = 1, 2, 3, ..., 10^5)
\end{eqnarray}

For example, one gets $\Phi(10, d) = 0.126$, when $\xi = 10$, and
$\Phi(100, d) = 0.184$, when $\xi = 100$.

\begin{figure}[!htbp]
\includegraphics[angle=0,scale=.31]{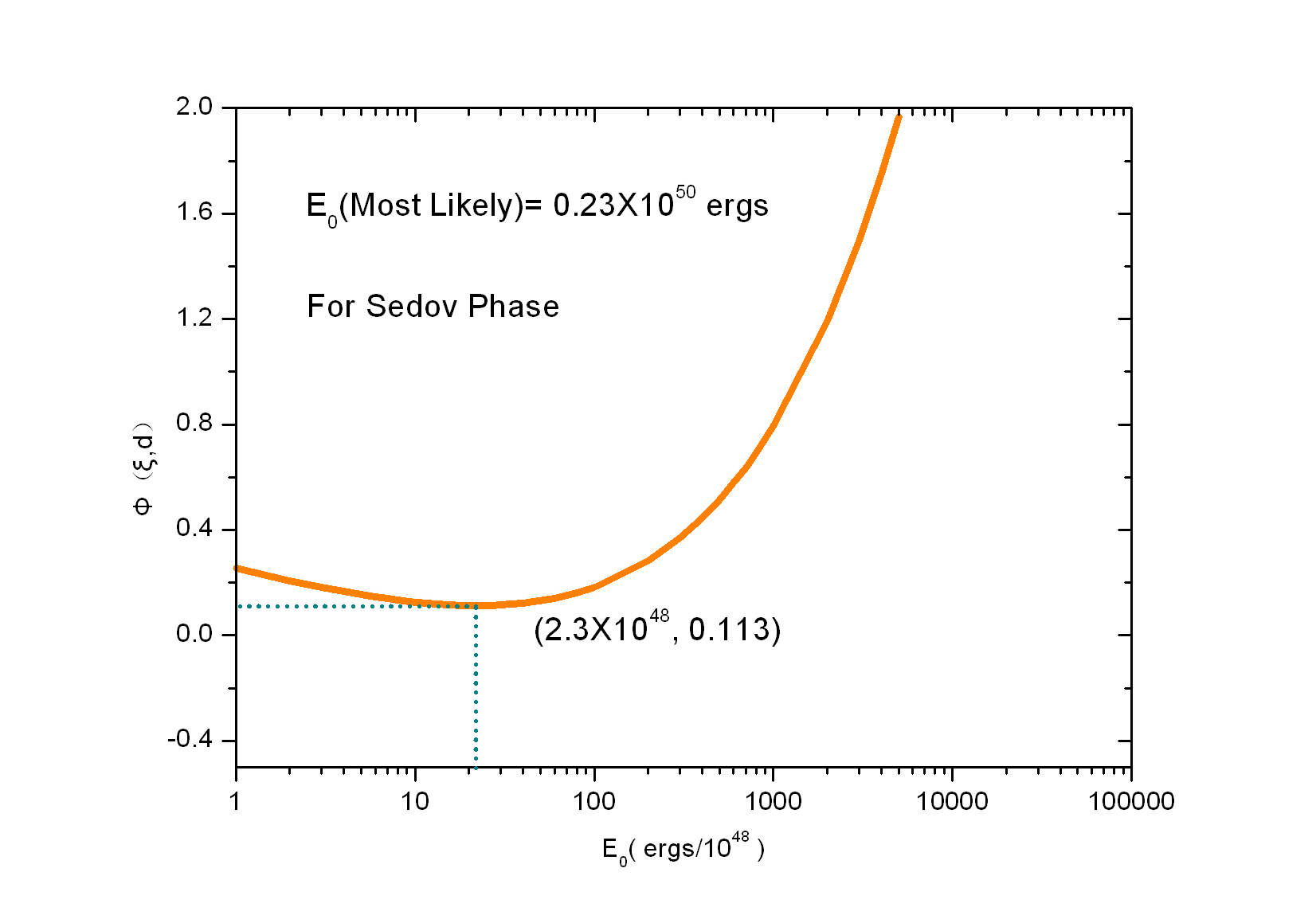}
\caption{The most likely value of supernova initial explosion energy
($E_0$) derived by comparison with already known distance ($d$) of
the shell-type remnants at Sedov-phase equals nearly to $0.23\times
10^{50}$~ergs. \label{gc2d}}
\end{figure}

Figure~\ref{gc2d} shows that the most likely value of supernova
initial explosion energy ($E_0$) derived by this method for the
shell-type remnants at Sedov-phase equals nearly to $0.23\times
10^{50}$~ergs.

There are some different methods to derive the distance value ($d$)
of supernova remnants (Xu et al. 2005) in Galaxy, and majority of
these distances are not taken advantage of the ($E_0$) value
obtained by other authors before. These already known distances were
derived by the optical proper motion/velocity, HI absorption,
association with CO/HI/HII region, HI column density, pulsar
parallax and optical velocity, or by various methods (e.g., Green
2004; Xu et al. 2005), but not by $\Sigma$-$D$ relation or Sedov
solutions. Therefore through comparison by distances to confirm SNe
explosion energy ($E_0$) is to some extent reasonable and causes
almost no self-contradiction.

\subsubsection{Comparison by ages}

Similarly we can compare these resolved ages ($t_{our}(\xi,i)$) above
with the already known ones ($t_{true}(i)$) in table~\ref{tab2nd} by
\begin{eqnarray}
\Phi(\xi, t) = \sum_{i=1}^{37} (t_{our}(\xi,i)-t_{true}(i))^2\nonumber\\
\div \sum_{i=1}^{n} t_{true}^2(i)\nonumber\\
(\xi = 1, 2, 3, ..., 10^5)
\end{eqnarray}

For example, one gets $\Phi(10, t) = 0.532$, when $\xi = 10$, and
$\Phi(100, t) = 0.507$, when $\xi = 100$.

\begin{figure}[!htbp]
\includegraphics[angle=0,scale=.31]{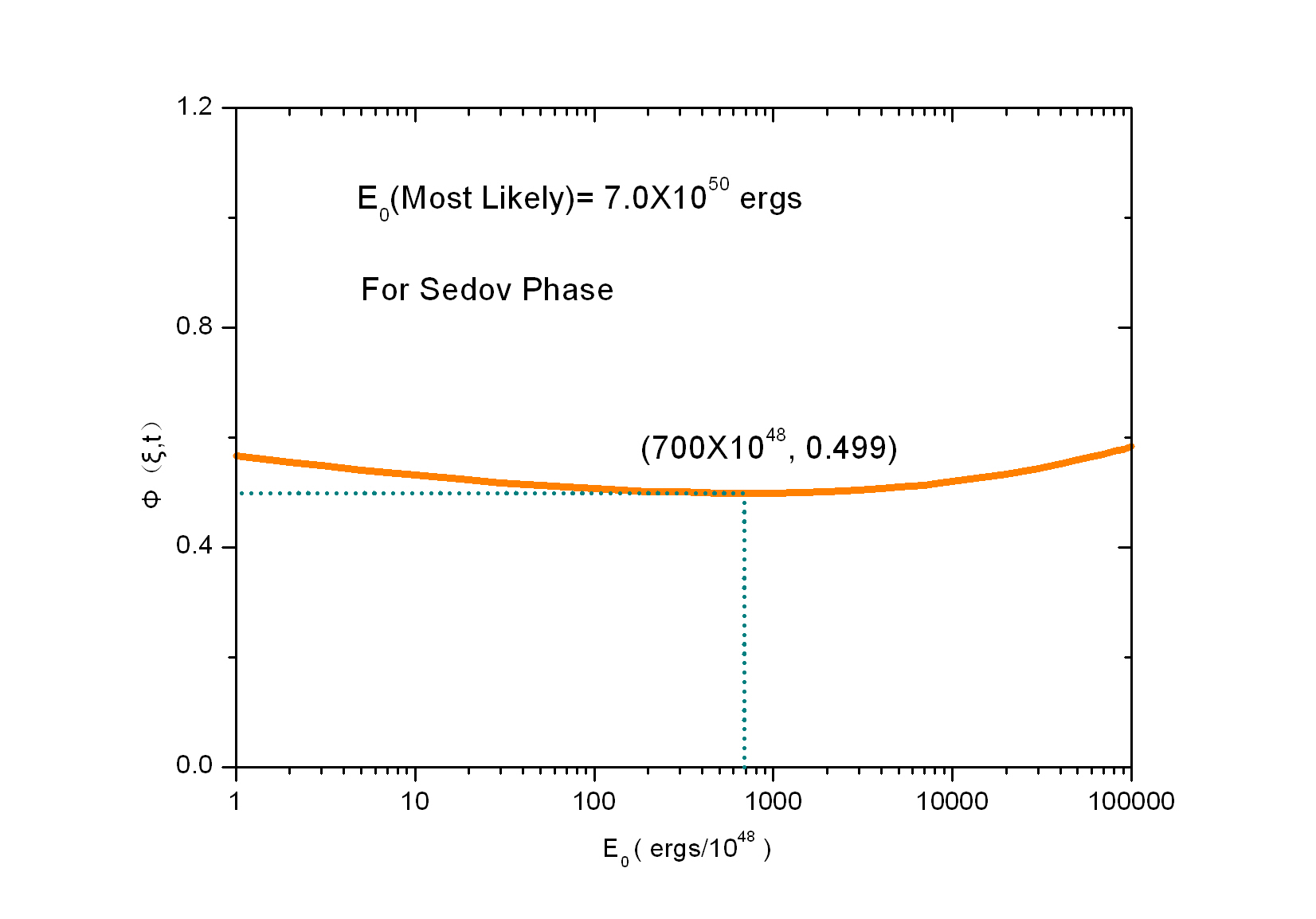}
\caption{The most likely value of supernova initial explosion energy
($E_0$) derived by comparison with already known age ($t$) of the
shell-type remnants at Sedov-phase equals nearly to $7.0\times
10^{50}$~ergs. \label{gc2t}}
\end{figure}

Figure~\ref{gc2t} shows that the most likely value of supernova
initial explosion energy ($E_0$) derived by this method for the
shell-type remnants at Sedov-phase equals nearly to $7.0\times
10^{50}$~ergs.

\begin{deluxetable}{lllllll}
\tablewidth{0pt} \tablecaption{List of the distance ($d$), Age ($t$)
and some other physical parameters of 20 shell-type Galactic SNRs of
which their diameter is larger than 36~pc. \label{tab3rd}}
\startdata
\hline\noalign{\smallskip}
\hline\noalign{\smallskip}
 Source & $t_{true}(i)$ & $d_{true}(i)$ & Dia.\tablenotemark{a} & size($\theta$) & $S_{1GHz}$ & Ref.\\
    $-$       &  yr      &   pc        &  pc    &  arcmin   &  Jy     &  $-$       \\
\hline\noalign{\smallskip}
   G8.7$-$0.1 &  15800   &   3900      &  51    &  45       &  80     &  G96       \\
  G18.8$+$0.3 &  16000   &   14000     &  57    &  17x11    &  33     &  D99,G04a  \\
  G49.2$-$0.7 &  30000   &   6000      &  52    &  30       &  160    &  KKS95, G04a\\
  G55.0$+$0.3 &  1100000 &   14000     &  71    &  20x15?   &  0.5    &  MWT98     \\
  G65.3$+$5.7 &  14000   &   1000      &  78    &  310x240  &  52     &  LRH80, R81\\
 G119.5$+$10.2&  24500   &   1400      &  37    &  90?      &  36     &  M00       \\
 G127.1$+$0.5 &  85000   &   5250      &  69    &  45       &  13     &  FRS84     \\
 G132.7$+$1.3 &  21000   &   2200      &  51    &  80       &  45     &  GTG80, G04a\\
 G156.2$+$5.7 &  26000   &   2000      &  64    &  110      &  5      &  RFA92     \\
 G160.9$+$2.6 &  7700    &   1000      &  38    &  140x120  &  110    &  LA95      \\
 G166.0$+$4.3 &  81000   &   4500      &  57    &  55x35    &  7      &  L89, KH91, G04a\\
 G166.2$+$2.5 &  150000  &   8000      &  186   &  90x70    &  11     &  RLV86     \\
 G182.4$+$4.3 &  3800    &   3000      &  44    &  50       &  1.2    &  KFR98     \\
 G205.5$+$0.5 &  50000   &   1600      &  102   &  220      &  160    &  CB99      \\
 G206.9$+$2.3 &  60000   &   7000      &  102   &  60x40    &  6      &  L86       \\
 G266.2$-$1.2 &  680     &   1500      &  52    &  120      &  50     &  K02, AIS99\\
 G296.5$+$10.0&  20000   &   2000      &  44    &  90x65    &  48     &  MLT88     \\
 G296.8$-$0.3 &  1600000 &   9600      &  47    &  20x14    &  9      &  GJ95, G04a\\
 G321.9$-$0.3 &  200000  &   9000      &  70    &  28       &  13     &  SFS89, S89\\
 G330.0$+$15.0&  $-$     &   1200      &  63    &  180?     &  350    &  K96       \\
\enddata
\tablenotetext{a}{Notes: Diameters were calculated by using the
distances together with the angular sizes in Green (2006)
catalogue.}
\end{deluxetable}

\begin{figure}[!htbp]
\includegraphics[angle=0,scale=.31]{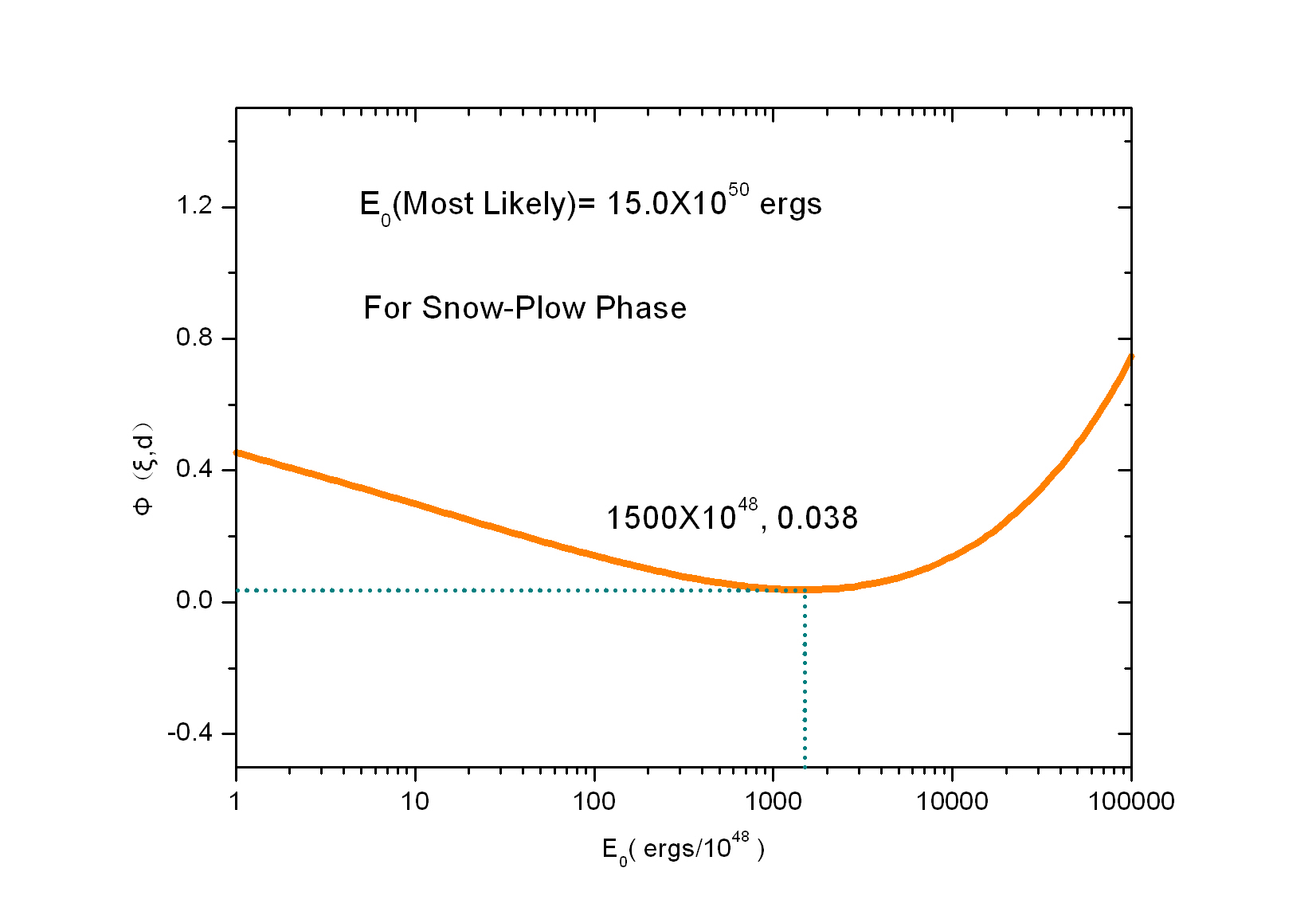}
\caption{The most likely value of supernova initial explosion energy
($E_0$) derived by comparison with already known distance ($d$) of
the shell-type remnants at Snowplow-phase equals nearly to
$15.0\times 10^{50}$~ergs. \label{gc3d}}
\end{figure}

\begin{figure}[!htbp]
\includegraphics[angle=0,scale=.31]{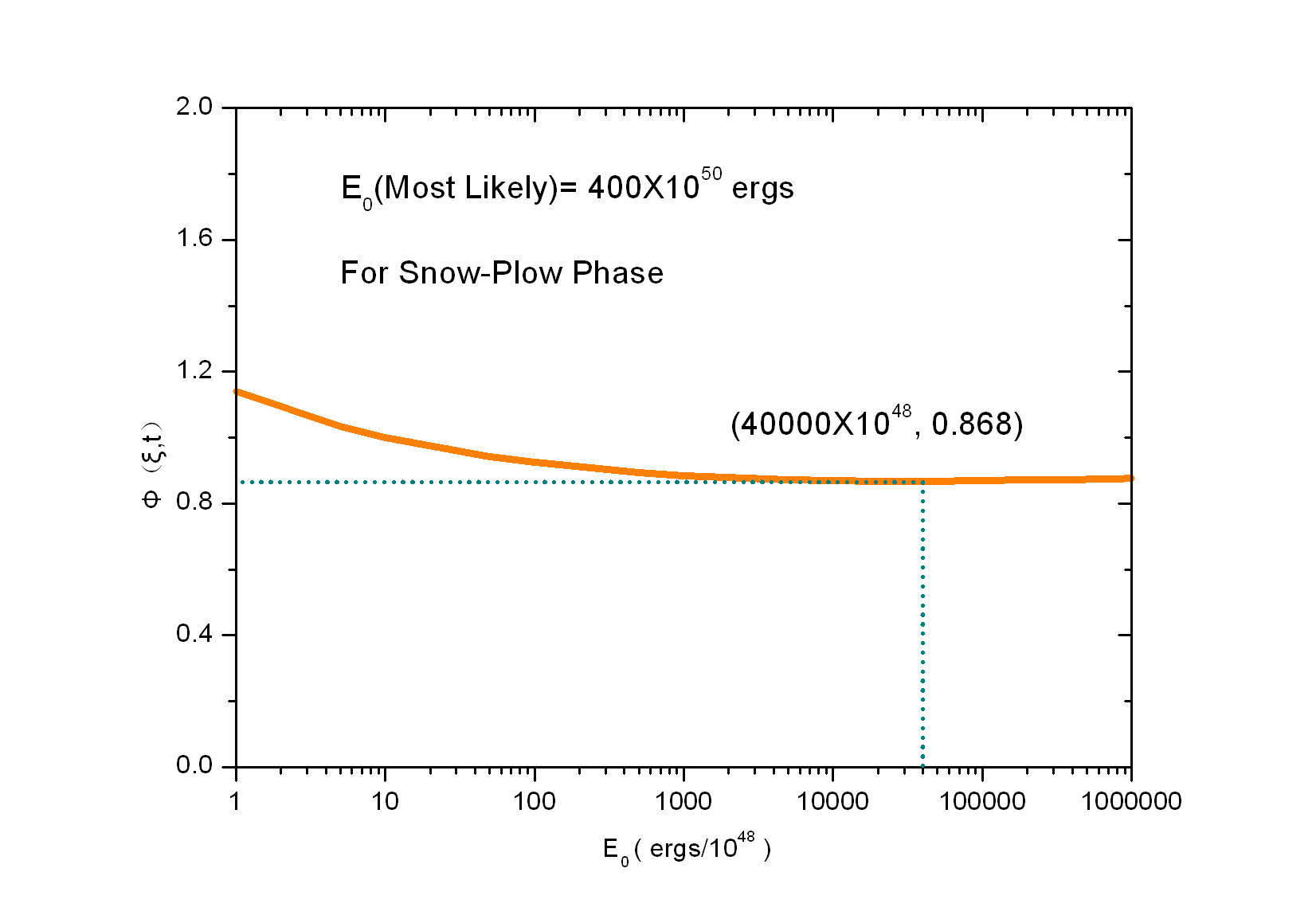}
\caption{The most likely value of supernova initial explosion energy
($E_0$) derived by comparison with already known age ($t$) of the
shell-type remnants at Snowplow-phase equals nearly to $400\times
10^{50}$~ergs. \label{gc3t}}
\end{figure}

There are many measures to obtain the age values ($t$) of remnants
(Xu et al. 2005), and a majority of these ages are not made use of
the $E_0$ value derived before. For example, if a remnant is
associated with a pulsar, we can estimate its age by using the
neutron star characteristic age obtained from the rotation period of
the pulsar (P) and the rate of change of period ($\dot{P}$) by
$t=P/2\dot{P}$ (Gotthelf et al. 2000). For SNRs with a known radius
($R$) and thermal temperature ($T$) taken from X-ray data, one can
obtain the age by $t=3.8\times10^2R_{\rm pc}(kT)^{-1/2}_{\rm
keV}$~yr (Seward et al. 1995). We can also calculate the SNR age by
$t~\approx~40000B^{-1.5}\nu_{\rm b}^{-0.5}$~yr, when a remnant has
its spectrum showing the usual break at frequency $\nu_b$ due to
synchrotron losses in a magnetic field $B$ (Bock et al. 2001).
Therefore through comparison by ages to determine SNe initial
kinetic energy ($E_0$) is somewhat meaningful and basically no
paradox.

But this measure to determine SNe energy $E_0$ through comparison by
ages is to some extent less reliable than that through comparison by
distances. Because the initial energy $E_0$ of some individual SNRs
derived from its age. One can see that the obtained remnants
distance is more independent to $E_0$ than the age be, and therefore
it causes less self-contradiction in our work.

\subsection{At radiative phase}

Similarly we list the group of equations for shell-type remnants at
the third stage (Koyama \& Meguro-Ku 1987, Kitayama \& Yoshida 2005,
Xu et al. 2005)

\begin{equation}
D_{pc} = 1.42 \left( \begin{array}{c}
\frac{E_0/10^{51}ergs}{n_{cm^{-3}}}
\end{array} \right) ^{5/21} t_{yr} ^{2/7}
\end{equation}

\begin{eqnarray}
\Sigma(D) = 1.505\times 10^{-19} \frac{S_{1GHz}}{\theta_{arcmin}^2}\nonumber\\
= 2.88\times 10^{-14} D_{pc}^{-3.8} n_{cm^{-3}}^2
\end{eqnarray}

\begin{equation}
t_{yr} = 10^5 n_{cm^{-3}}^{-3/4} \left( \begin{array}{c}
\frac{E_0}{10^{51}ergs} \end{array} \right)^{1/8}
\end{equation}

Here, $D_{pc}$, $t_{yr}$, $n_{cm^{-3}}$, $S_{1GHz}$ and
$\theta_{arcmin}$ is defined as in section~\ref{sedov} as well as
their units, and $tan \left(\begin{array}{c}
\frac{\theta_{arcmin}}{2} \end{array} \right) =
\frac{D_{pc}}{2d_{pc}}$. The fluxes $S_{1GHz}$ and observational
angle $\theta_{arcmin}$ are already known to us for each remnant
(table~\ref{tab3rd}) of which we regard these remnants evolving at
the snow-plough phase since their diameter larger than 36~pc. When
the initial energy $E_0 = \xi \times E_{48}$ assumed, then the
remnant diameter $D_{pc}$ (and distance $d_{pc}$), age $t_{yr}$ and
electron density $n_{cm^{-3}}$ can be obtained.

One can see the equations (6) and (8) are rather different from
equations (1) and (3). But formulae (7) and (2) are completely the
same.

For a certain supernova remnant ($i$) ($i = 1, 2, 3, ..., 20$)
(table~\ref{tab3rd}) and assumed $E_0 = \xi \times E_{48} (\xi = 1,
2, 3, ..., 10^5)$ in units of $10^{48}$~ergs, we can get the remnant
distance $d_{our}(\xi,i) (= d_{pc})$ and the age $t_{our}(\xi,i) (=
t_{yr})$ by solving the equations group above. Then we compare them
with the already known parameters $d_{true}(i)$ and $t_{true}(i)$
($i = 1, 2, 3, ..., 37$) listed in table~\ref{tab3rd} in order to
derive the most likely original energy of SNRs.

\subsubsection{Comparison by distances}

We can compare these resolved distances ($d_{our}(\xi,i)$) above with the already
known ones ($d_{true}(i)$) listed in table~\ref{tab3rd} by
\begin{eqnarray}
\Phi(\xi, d) = \sum_{i=1}^{20} (d_{our}(\xi,i)-d_{true}(i))^2\nonumber\\
\div \sum_{i=1}^{n} d_{true}^2(i)\nonumber\\
(\xi = 1, 2, 3, ..., 10^5)
\end{eqnarray}

Figure~\ref{gc3d} shows that the most likely value of supernova
initial kinetic explosion energy ($E_0$) derived by this method for
the S-type remnants at Snowplow-phase equals nearly to $15.0\times
10^{50}$~ergs.

\subsubsection{Comparison by ages}

Similarly we can compare these resolved ages ($t_{our}(\xi,i)$) above with the
already known ones ($t_{true}(i)$) in table~\ref{tab3rd} by
\begin{eqnarray}
\Phi(\xi, t) = \sum_{i=1}^{20} (t_{our}(\xi,i)-t_{true}(i))^2\nonumber\\
\div \sum_{i=1}^{n} t_{true}^2(i)\nonumber\\
(\xi = 1, 2, 3, ..., 10^5)
\end{eqnarray}

Figure~\ref{gc3t} shows that the most likely value of supernova
initial explosion energy ($E_0$) derived by this way for the
shell-type remnants at Snowplow-phase equals nearly to $400\times
10^{50}$~ergs.

\subsection{Final results}

From Fig.~\ref{gc2d} to Fig.~\ref{gc3t}, we get the least value of
$log_{10}E_0$ instead of $E_0$, therefore one has
\begin{eqnarray}
log_{10}E_0 = log_{10}E_{02d} + log_{10}E_{02t}\nonumber\\
+log_{10}E_{03d}+log_{10}E_{03t}
\end{eqnarray}
Here, $E_{02d}$, $E_{02t}$, $E_{03d}$ and $E_{03t}$ corresponding to
the 4 $E_0$ values in from Fig.~\ref{gc2d} to Fig.~\ref{gc3t}.
$\-E_0$ is the typical explosion energy of shell type remnants.

Thus we have $\-E_0 = 0.99 \times 10^{51}$~ergs.

The publicly accepted value of the SNe initial kinetic energy is $1
\times 10^{51}$~ergs.

\section{Discussion and Summary}

To combine theoretical results together with observational ones, we
have derived the supernova average initial explosion energy. Such
value is obviously not the same as that to a individual remnant. For
individual SNR the initial energy is various for different SN
explosion events, but it might be near an approximation. In
Fig.~\ref{distrib} the distribution of the initial energies of 44
SNRs range from $10^{49}$~ergs to $10^{52}$~ergs, mainly concentrate
to about $10^{50} \sim 10^{51}$~ergs. Our 4 theoretical outcomes
range from $0.23 \times 10^{50}$~ergs at minimum to $400 \times
10^{50}$~ergs at maximum. It seems that they are to some extent in
good consistency regarding their divergence. The relative divergence
of explosion energies about $\frac{10^3}{10^{51}}$ is a rather small
number and acceptable.

From table~\ref{tabdistr} one can directly calculate the mean
initial kinetic energy $\-E_0 = 1.7 \times 10^{51}$~ergs. Or $\-E_0
= 0.85 \times 10^{51}$~ergs after excluding SNR G0.0+0.0, namely Sgr
A East which owns an extremely large initial energy value. These is
also another method to obtain the initial energy by the group of
equations combining the physical parameters listed in
table~\ref{tab2nd} and table~\ref{tab3rd}, when we regard the age
and distance values as the already known ones. We gain $E_0 (i)$ ($i
= 1, 2, 3, ..., 37$) or ($i = 1, 2, 3, ..., 20$), and the average
$E_0$ is then derived. Here, in the paper our special method is
different from this, and has provided one more measure to derive the
SNe typical energy by taking somewhat more numbers of the SNRs with
intensified credit. Furthermore, many of the supernova remnants in
table \ref{tab2nd} or table \ref{tab3rd} are not the same as those
in table \ref{tabdistr}. Namely the explosion energy $E_0$ of rather
some remnants in table \ref{tab2nd} and table \ref{tab3rd} are
unknown to us. Because for most of the individual supernova remnant
in Galaxy their explosion energy is rather difficult to derive, here
our statistical $E_0$ value obtained can be used to compute other
physical parameters, i.e. the distance, age, ISM electron density,
magnetic field, etc. of which the errors resulted in will be to some
extent rather small.

Here we did not distinguish different type supernova of our adopted
sample in the numerical analysis. For Type-I and Type-II SN the
initial explosion energy might be not similar, but one can expect
that this deviation will be somewhat small.

The average value of the most likely initial explosion energy of
supernova remnants thus equals to about $0.99\times10^{51}$~ergs in nice
consistence with the accepted value.

\acknowledgments

JWX thanks J.S. Deng and Y.Z. Ma for their assistance and help
during the paper work.

\end{document}